\documentclass[submission,copyright,creativecommons]{eptcs}
\providecommand{\event}{AiSoS 2013} 
\usepackage{breakurl}             
\usepackage[utf8]{inputenc}
\usepackage{amsmath,amsthm,amsfonts,amssymb}
\usepackage{xcolor}
\usepackage[nounderscore]{syntax}
\usepackage{graphicx}
\usepackage{enumitem}

\newtheorem{definition}{Definition}

\newtheorem{proposition}{Proposition}
\newtheorem{example}{Example}

\newcommand{\D}{\mathbb{D}}
\newcommand{\kw}[1]{{\bf #1}}

\title{Contracts and Behavioral Patterns for SoS: The EU IP DANSE approach}

\author{
Alexandre {\sc Arnold}
\institute{EADS Innovation Works\\ Toulouse, France}
\email{\quad Alexandre.Arnold@eads.net}
\and
Benoît {\sc Boyer}, Axel {\sc Legay}
\institute{INRIA - Rennes Bretagne Atlantique\\ Rennes, France}
\email{First.Last@inria.fr}
}
\def\titlerunning{Contracts and Behavioral Patterns for SoS}
\def\authorrunning{A. Arnold, B. Boyer, A. Legay}
\begin{document}
\maketitle

\begin{abstract}
  This paper presents some of the results of the first year of {\sc Danse},
  one of the first EU IP projects dedicated to SoS. Concretely, we
  offer a tool chain that allows to specify SoS and SoS requirements
  at high level, and analyse them using powerful toolsets coming from
  the formal verification area. At the high level, we use UPDM, the
  system model provided by the british army as well as a new type of
  contract based on behavioral patterns. At low level, we rely on a
  powerful simulation toolset combined with recent advances from the
  area of statistical model checking. The approach has been applied to
  a case study developed at EADS Innovation Works.
\end{abstract}

\section{Introduction}
\label{sec:introduction}
While SysML~\cite{SysML}, the Systems Modeling Language derived from UML~\cite{UML}, has been
widely adopted for Systems Engineering applications, the specificities
of Systems of Systems (SoS) fostered the creation of further
customizations. The Unified Profile for DoDAF and MoDAF (UPDM)~\cite{UPDM}, based
on the US and UK military architectural framework, is one of them and
is used on a regular basis in SoS Engineering.

Specific extensions of SysML/UPDM are considered in {\sc Danse}~\cite{project-DANSE},
one of the first European project aiming at developing a methodological and technical
framework for SoS Engineering with associated tool support. This
framework shall support the SoS architect from the modeling activities
to the analysis phase (abstraction, simulation, formal verification),
especially by providing concrete solutions to address common SoS
issues: constant evolution of a large-scale SoS and its stakeholders'
needs, unexpected emergent behaviors, limited awareness of the global
situation...

In the frame of {\sc Danse}, we extend the language of SysML/UPDM to
add formalized requirements for an SoS. Formalizing the SoS goals
makes it possible to verify them automatically (with an adjustable
probability) using a statistical model checker such as Plasma-Lab~\cite{Plasma-Lab,JegourelLS12} in
combination with a simulation platform such as DESYRE. The challenge
is to propose a high-level formal language that is directly usable by
an SoS architect, while being still automatically translatable to the
expressive low-level specification of the model checker, in a similar
way to editors like IBM Rhapsody that could make an executable
specification (FMI) out of a high-level formalism (SysML/UPDM
behavioral diagram).

For our purpose, the low-level specification is the Bounded Linear
Temporal Logic (B-LTL), an extension of the Linear Temporal Logic in
which each temporal operator is bound by a temporal constant. This
logic is expressive enough to cover a large set of properties and to
write static as well as behavioral SoS goals. But this logic is
defined using the standard temporal operators, which are quite
low-level: defining complex properties often requires to interlock
several layers of nested operators. Writing or understanding such
formulas is difficult, thus error-prone, and does not fit at all with
our target of a clear and simple specification language.

So we propose in this paper the very first contract language for
SysML/UPDM, defined using a strong B-LTL based semantics, but close to
hand written English requirements for SoS on the surface. This
language of goal formalisation, which is developed in the scope of the
{\sc Danse} project, is called the Goal and Contract Specification
Language (GCSL).

GCSL makes use of the Object Constraint Language (OCL), a formal
language by the Object Management Group (OMG)~\cite{OMG} used to
describe static properties on UML models, thus also on SysML/UPDM
ones. OCL can be used for a number of different purposes, but
especially as a model-based query language and for writing
expressions, which perfectly suits our needs here. GCSL also reuses
the Contract Specification Language (CSL)~\cite{patterns-SPEEDS}, developed in the previous
SPEEDS European project~\cite{project-SPEEDS}, which comes with convenient temporal
patterns.
The three key elements required for the formalization of behavioral
goals and the way we address them in our approach are (1) being able to refer to model elements: use of the same names as in the SysML/UPDM model, (2) being able to write static properties about them: use of OCL and 
(3) being able to integrate these expressions inside behavioral patterns: use of CSL patterns.

After a short description of the SoS modeling in
Section~\ref{sec:semantic}, this paper presents in
Section~\ref{sec:contr-lang-updmlsysm} the GCSL based on the semantics
of UPML modeling, thus we show how to translate the properties into
B-LTL formulas (Section~\ref{sec:translation}) into order to check
them using the statistical model checking framework for SoS
(Section~\ref{sec:stat-model-check}). Finally, Section~\ref{sec:illustration} illustrates
the approach applied to the case study of {\sc Danse}.




\section{System of Systems Modeling}
\label{sec:semantic}

\subsection*{Overview of SysML/UPDM}
SysML~\cite{SysML} is a general-purpose modeling language defined as an extension
of a subset of the Unified Modeling Language (UML)~\cite{UML} using UML's profile
mechanism. SysML is used for Systems Engineering applications, whereas
UML is more targeted towards object-oriented Software Engineering. A
large set of diagrams is provided with SysML to model a system's
requirements, structure (e.g. block definition diagram, internal block
diagram), behavior (e.g. state machine, activity diagram), etc.

Using the same UML's profile mechanism, another language built on top
of UML/SysML has de facto become a standard for SoS architects:
UPDM. This profile is the result of the unification effort of the US
Department of Defense and the UK Ministry of Defense architecture
frameworks and associated meta-models. It adds a layer of new
meta-objects that are typically (but not exclusively) used in the
context of military SoS, as well as a significant amount of predefined
views (e.g. system views, operational views, capability views) which
help splitting the whole modeling activity in smaller tasks.

The executable part of a UPDM modeling can be compiled into a program
based on the Functional MockUp Interface (FMI)~\cite{FMI} that defines
a standardized interface used in simulations of complex systems. In
{\sc Danse}, the SoS is compiled into FMI program and executed by the
simulation engine DESYRE~\cite{Ales}.  This whole FMI program can be
considered as a state transition system, e.g. the formal semantics on
which we will use to propose our language.  The states denote the
global states of the SoS, e.g. the result of collecting the internal
states of each constituent in the SoS. The transitions denote the
actions and events that occur in the SoS and eventually modify the
internal state of some system constituents and thus, the global state
of the system.

\begin{definition}[State Transition Systems]
Let $X$ be a set of variables that are mapped to the values of $\D$,
the set of all possible values. We define $S$ a set of states. Each
state $s$ is characterized by a mapping $\mu_s: X \to \D$ such that the
valuation of any variable $y\in X$ in the state $s$ is $\mu_s(y)$. We thus
define a transition system as a 4-tuple $<S, s_0, R,\{\mu_s\ |\ s \in
S\}>$ such that
\begin{itemize}
\item $s_0 \in S$ contains the initial states of the system
\item $R \subseteq S \times S$ is the transition relation. We use the
  more convenient notation $s \to s'$ to denote $(s, s') \in R$.
\end{itemize}

All valid execution (or run) of a transition system is a sequence of
states led by the $R$ from any initial state. A run of length $n$ will
denoted as $\pi = s_0; s_1; s_2; \dots; s_n$ where $s_0 \in I$ and
$s_i \to s_{i + 1}$ holds for $0 \le i < n$. Each transition system has
a global clock, which is denoted by the variable $t$. We note $t_i = \mu_{s_i}(t)$,
the observed time value of $t$ when the executed system reaches the state $s_i$.
For any execution path the system is in state $s_i$ when $t_i \le t < t_{i+1}$ and
the evolution of the time is monotonically increasing, e.g. $t_i < t_{i+1}$.
\end{definition}

In the first year of the {\sc Danse} project we limit ourselves to systems
of systems who environment's behaviors are fully known in advance
(hence representable via state transition systems), like it is the
case for most of adaptive systems studied in the
litterature~\cite{ZC06,Ghezzi11,ChengLGIMABBBCSDFGGGKKKLMMMPSTTWW09,FHNPSV11}.
The reason is that this corresponds to the current possibilities of
the UPDM. In future work, we will study more complex aspects such as
unknown environment, hence more complex dynamicity features. For this,
we will first have to consider extension of the UPDM model.

\subsection*{Stochastic aspects of the model}
Stochastic modeling is a way to describe behaviors that are not
deterministic by nature, or to abstract a behavior that is simply too
complex to be modeled explicitly (as white box). So it is typically
very useful in a SoS context. Behavioural modeling in SoS examples
such as an Emergency Response to a city fire typically shows numerous
attributes/parameters that would not be deterministic, such as the
time between two fires or the duration of an action performed by a
human.

A first proposal of how to put stochastic data in the SysML/UPDM model
has been integrated into the {\sc Danse} project. It is based on a set
of attribute stereotypes that can be applied to any block
attribute. This idea is close to the suggestion of the non-normative
distribution extensions made in appendix of the SysML 1.3
specification, but adds the possibility to regenerate a
distribution-based random value whenever needed (and not only at
initialization). This addition is important because even the same
person does never perform the same task in the exact same amount of
time, so that the duration of the task shall be recalculated every
time.

Adding stochastic data to the SoS model implies of course that each
simulation is likely to generate a different trace than the previous
ones, and as a consequence that one run will not be enough to verify
whether the SoS meets its requirements or not. This is why being able
to automate this verification process in a mathematical way (provided
the requirements are formalised) is a great support for the SoS
architect when assessing a candidate architecture.

Since the SoS we consider exhibit some stochastic behaviors, each run
has an associated probability of being executed. This probability is
given by an unkonwn distribution due to the high complexity of the
model: a system is designed by the paralell composition of components
that may have a stochastic behavior.



\section{A Contract Language for UPDM/SysML Requirements}
\label{sec:contr-lang-updmlsysm}

Before defining the new contract language, we introduce the notion of contract
for the SoS formalized as a stochastic state transition systems.

\begin{definition}[Contracts for State Transition Systems]
  \label{def:contract}
  A contract is defined as a pair $(A, P)$ where $A$ and $P$ are
  respectively called the Assumption and the Promise. Considering a
  state transition system, $A$ and $P$ are properties about the execution
  of the system. Thus, a contract for the system specifies what the
  system shall ensure the promise when the system shall satisfy the
  assumption. The notation $Sys \models (A, P)$ means that the
  contract $(A, P)$ is satisfied by the system $Sys$. Relying on the
  state transition system semantics, the satisfaction of a contract is
  \[Sys \models (A, P) \quad \textrm{iff} \quad \forall \pi,\ \pi
  \models A \Rightarrow \pi \models P\] where $\pi$ is a valid run of
  $Sys$ and $\pi \models A$ (or $P$) means the run $\pi$ satisfies the
  assumption $A$ (the promise $P$ resp.).
\end{definition}

For stochastic systems, it is generally more meaningful to
quantify how a system satisfies a contract: this valuation is
given by the probability that the system satisfies the
contract. Intuitively, if the distribution to execute each run of a
given stochastic system is known, the probability that this system
satisfies the contract is the sum of the probabilities of all the runs
that satisfy the contract (see Section~\ref{sec:stat-model-check}).

\begin{definition}[Contracts for Stochactic State Transition Systems]
  \label{def:stoch-contract}
  Let be a stochastic system $Sys$, a contract $(A, G)$ and a
  threshold value $k \in [0..1]$. For the system $Sys$, we now
  consider the contract $P_{\sim k}(A, G)$, where $\sim \in \{<, \le,
  =, \ge, >\}$ and $0 \le k \le 1$. The contract is satisfied if and
  only if the relation holds, e.g. if the probability $p$ of $Sys
  \models (A, G)$ satisfies the relation $p \sim k$.
\end{definition}

In this work, for efficiency reasons, we decided to estimate the
probability $p$ using statistical model checking rather than computing
it with a numerical approach such as Prism~\cite{PRISM}.  Another
reason to use SMC is that it relies on monitoring traces, hence it
allows to verify properties that cannot be expressed in classical
logics. In this paper, this aspect will not be explored, but it is a
main topic of {\sc Danse}. SMC consists in verifying the property
(here contract) against several simulations of the system. Then, an
algorithm from the statistic area is used to estimate the probability
to satisfy the property. The contract to monitor is translated into a
B-LTL formula (see Section~\ref{sec:translation}) that characterizes a
set of simulation traces. Thus, the simulation monitoring consists of
observing each simulation to decide if the B-LTL formulas holds or
not.



We now introduce the language to express the assumptions and promises
dedicated to the System of Systems.  The GCSL syntax for patterns is a
combination of the Object Constraint Language (OCL) and the contract
patterns of the CSL à la "SPEEDS"~\cite{project-SPEEDS}.  The SPEEDS
contract specification patterns are introduced in the SPEEDS
Deliverable D.2.5.4 "Contract Specification Language
(CSL)"~\cite{patterns-SPEEDS} and used to give a high-level
specification of real-time components. They have been introduced to
enable the user to reason about event triggering that are equivalently
replaced in {\sc Danse} by property satisfaction. The properties handled by
these patterns are about the state of a SoS. We use OCL to specify
these state properties. This language allows to build some behavioral
properties to express some temporal relations about facts or events of
the system denoted by the state properties. It is sufficiently
powerful to describe precisely a state of a SoS.  Here, we will only
consider a subset of the OCL language, but it is not unrealistic to
consider a larger subset of OCL to describe the requirements. We
restrict the language here to express some properties that can be
verified using the SMC techniques applied to SoS's.

We briefly recall the notion of Collection that we will use in the
rest of the paper.
\paragraph{Collections in OCL:} in OCL, it is the usual way to define
some properties about set of elements in a system. Considering a SoS
as a state transition system, the root identifier {\tt SoS} denotes
$\sigma$ the state currently reached by the SoS.  The collections
allow to handle some set of instances of components in the current
state $\sigma$. A collection built over the state $\sigma$ can be
viewed as a projection of $\sigma$: it is defined by selecting some
component instances or attribute values in the state $\sigma$.

For example, the expression ${\tt SoS.itsFireStations}$ denotes
collection of all the instances of type {\tt FireStation} at state
$\sigma$. OCL defines some operators that can be applied to any
collection: ${\tt SoS.}$ ${\tt itsFireStations\rightarrow size()}$
counts the number of instances of type {\tt FireStation}.  The most
important feature of the collection is the predicates we can define
using quantification:
\begin{itemize}
\item ${\tt SoS.coll\rightarrow forAll(x| \phi(x))}$ denotes that for
  all element {\tt x}, which belongs to the collection {\tt SoS.coll},
  the property $\phi(x)$ holds.
\item ${\tt SoS.coll\rightarrow exists(x| \phi(x))}$ denotes for that
  there exists one element {\tt x}, which belongs to the collection
  {\tt SoS.coll}, the property $\phi(x)$ holds.
\end{itemize}

\subsection*{State properties in OCL}
Originaly introduced to supplement UML, the Object Constraint Language
(OCL)~\cite{OCL} is particularly adapted to describe the internal
state of a component. The Object Constraint Language is a rather
simple-to-write, yet formal text language that provides constraint and
object query expressions based on any meta-model, so for instance the
SysML/UPDM ones. It has a concise notation for accessing, collecting,
filtering and evaluating model elements. More generally, it allows to
write invariants on a model, that we use in our approach to write the
static properties that we insert in the behavioral contracts. As we
will see in the following paragraphs, we also pushed the concept
further by sometimes embedding a CSL pattern inside an OCL-like
expression, when we want to state that the pattern shall hold for some
or all elements in a set. We recall some OCL notations used in the
rest of the paper, but the reader can find the whole specification
in~\cite{OCL}.  Components store internal values into attributes that
are denoted by the standard dot-syntax. For example, the number of
people in the district 1 at $\sigma$, the state reached by the SoS, is ${\tt
  district_1.population}$. More particular to OCL, it is also possible
to define a collection of attributes using the same syntax: the
expression ${\tt SoS.itsDistricts.population\rightarrow sum()}$
denotes the number of total people. For the sake of clarity in the
rest of the paper, we only focus on the {\tt Collection} type without
considering all its refinements ({\tt Set}, {\tt Ordered Sets},
\dots), and the subset of Boolean and arithmetic expressions over the
attributes of the SoS' component instances.

\subsection*{The behavioral patterns}
The semantics of the patterns is based on the satisfiability of any
predicate on the whole set of execution paths that defines the pattern,
which the definition of the following patterns are based
upon. Consider the state property $\Psi$ and a time value sequence
$t_0, t_1, \dots, t_n$ that defines the state sequence $\sigma_0,
\sigma_1, \dots, \sigma_n$ such that $t_i$ is the time value where the
system reaches $\sigma_i$. In other words, the system is in state
$\sigma_i$ when $t_i \le time < t_{i+1}$.

\begin{figure}[h!]
  \centering
  \includegraphics[width=10.5cm]{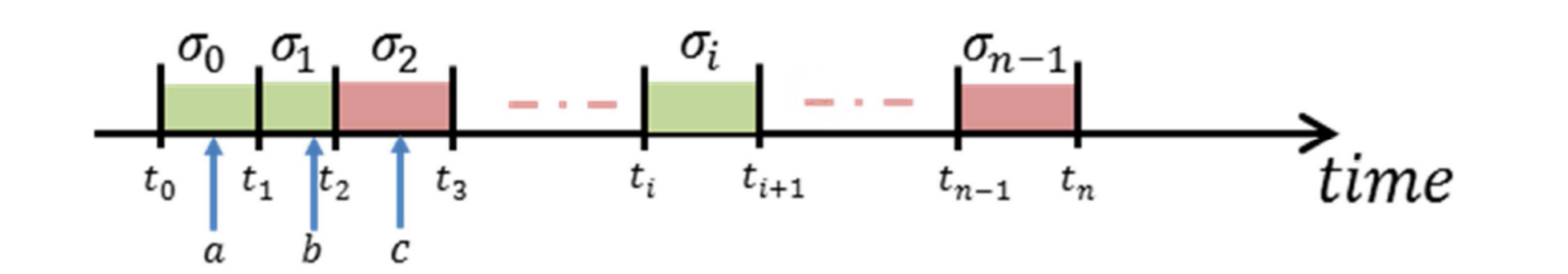}
  \caption{Satisfaction of $\Psi$ during an execution path.}
  \label{fig:satisfaction}
\end{figure}

Figure~\ref{fig:satisfaction} illustrates the satisfaction of a
state property $\Psi$, e.g. the green state $\sigma_0$, $\sigma_1$ and
$\sigma_i$ are the only states of the sequence that satisfy $\Psi$. It
means that $\Psi$ holds when $time \in [t_0, t_2) \cup [t_i,
t_{i+1})$. We observe that $\Psi$ holds continuously for $\sigma_0$,
$\sigma_1$, hence the number of occurrences where $\Psi$ holds is 2
during the time interval $[t_0, t_n)$. If we finally consider any the
time ticks $a$, $b$ and $c$, $\Psi$ holds during $[a,b]$ but does not
during $[a,b]$ nor $[b,c]$ and the occurrence number of $\Psi$ is $1$
in $[a,b]$, $[a,c]$ or $[b,c]$.

We define some selected patterns, but the more exhaustive list can be found in Appendix~\ref{sec:patterns}.
These patterns proved very useful for SoS applications. We assume that $\Psi$ and $\Psi_i$ are state properties
and $a, b, c$ are time constants such that the time intervals defined in the patterns are valid.
\begin{description}
\item \kw{whenever} $\Psi_1$ \kw{occurs} $\Psi_2$ \kw{does not occur during following} $[a,b]$
    \begin{center}
    \includegraphics[width=.5\linewidth]{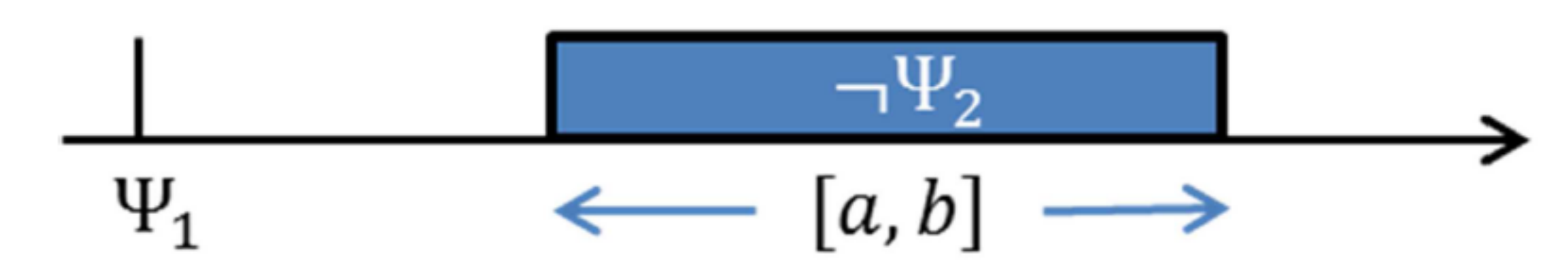}
  \end{center}
  This pattern specifies that $\Psi_2$ is never satisfied during the
  relative interval $[a, b]$ after $\Psi_1$, i.e. $\neg\Psi_2$ holds during $[a, b]$.
  By relative we means that when $\Psi$ occurs at $t$, the relative interval corresponds
  to $[a+t, b+t]$.

\item \kw{Whenever} $\Psi_1$ \kw{occurs} $\Psi_2$ \kw{occurs within} $[a,b]$
  \begin{center}
    \includegraphics[width=.5\linewidth]{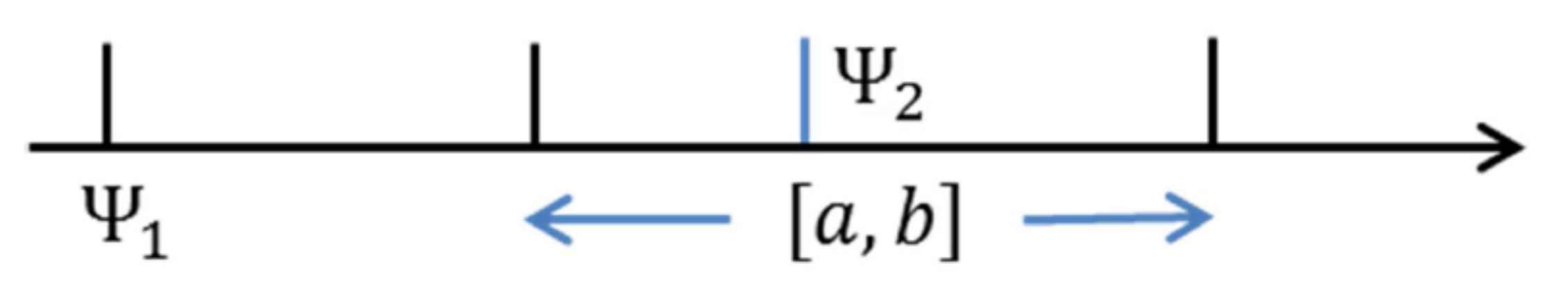}
  \end{center}
  The constraint $\Psi_2$ must be satisfied at least once during $[a,b]$ after $\Psi_1$.

\item
 $\Psi$ \kw{during} $[a,b]$ \kw{implies} $\Psi_1$ \kw{during} $[a,c]$ \kw{then} $\Psi_2$ \kw{during} $[c, b]$
 \begin{center}
   \includegraphics[width=.5\linewidth]{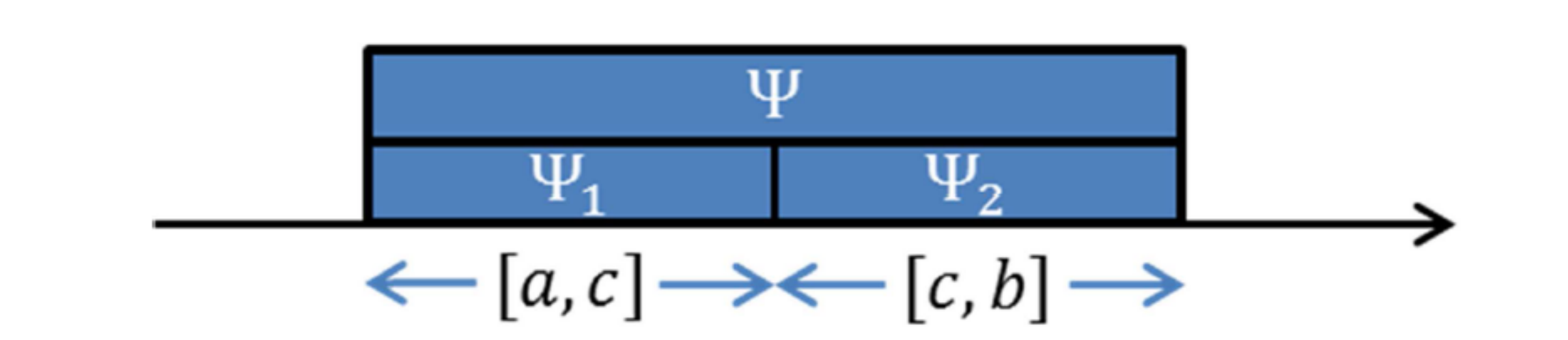}
 \end{center}
 Whenever $\Psi$ holds during $[a, b]$ there exists a split at $c$ of $[a, b]$ such that $\Psi_1$ holds during $[a,c]$ then $\Psi_2$ holds during $[c,b]$.
\end{description}

The CSL patterns are originally designed to specify the behavior of
any component instance by totally abstracting its environment without
quantification. It is not possible to specify a contract about the
interaction between two anonymous components. By anonymous, we mean
that no particular instance is explicitly referenced by the component
identifier. Let us consider a SoS with a set of components {\tt District}
and two {\tt District} properties $Psi_1$ and $\Psi_2$ in OCL.
The patterns allow to express the behavioral property for some explicit
component, e.g. \kw{Whenever} $[\Psi_1({\tt district_1})]$ \kw{occurs} $[\Psi_2({\tt district_1})]$ \kw{occurs within} $[a,b]$, it is not possible to generalize the behavioral property to any {\tt District} of the system, e.g. 
a property like "For all {\tt district}, \kw{Whenever} $[\Psi_1({\tt district})]$ \kw{occurs} $[\Psi_2({\tt district})]$ \kw{occurs within} $[a,b]$".

To overcome this important limitation, we extend the proposed grammar (see Appendix~\ref{sec:patterns})
by overlapping the patterns with the OCL collection predicates, e.g. {\tt forAll(x|\dots)} and {\tt exists(y|\dots)}.
Then, the generalized behavioral property presented below is now:

\[\begin{array}{ll}
  {\tt SoS.itsDistricts\rightarrow forAll(district\ |}\\
\hspace*{4cm}{\tt \kw{Whenever}\ [\Psi_1({\tt district})]\ \kw{occurs}\ [\Psi_2({\tt district})]\ \kw{occurs\ within}}\ [a,b])

\end{array}\]

The root collection {\tt SoS.itsDistricts} is defined on the initial state $\sigma_0$ of the SoS.
In SoS, the initial state is describe by the Internal Block Diagram that is defining the initial 
state of each component. Using these OCL predicates for quantify the
patterns keep the language not so different in comparison with the
original OCL, except we restrict the nesting capability. The OCL syntax allows to nest the quantification
without any limit.  If there is no theoretical reasons to have limit,
we impose a limit of 2 nested quantifications in our language.
From the verification point of view, a behavioral formula with more nested quantifications is not
practically check-able. Moreover, we never need more to express the
requirements of CEA incubator in {\sc Danse}.  So we assume in the
next, that the patterns have are of the form ${\tt SoS.coll_1\rightarrow
  forAll(x | SoS.coll_2\rightarrow forAll(y | \dots \kw{Pattern}(x, y) \dots)}$, where \kw{Pattern} is any behavioral pattern.

Another important limitation of this combination OCL + patterns
is the inability of express property about cumulative values during an execution path: to solve
this problem we introduce the path operators ${\tt mean}()$, ${\tt sum}()$, ${\tt prod}()$ to denote
the value of a numerical expression: for example, ${\tt mean(district_1.}$ ${\tt population)}$ denotes the
average value of the attribute ${\tt district_1.population)}$ computed with the values obtained of the different state of the path.

\subsection*{Examples of Requirements}
Table~\ref{tab:requirements_CAEincubator} illustrates the kind of
properties that we will express with our language. We use syntactic
coloring to distinguish the different parts of the language used in
the property: the words in red are identifiers from the model, the
blue part is from OCL and bold black keywords are temporal
operators. These requirements show the capabilities of our language
using different requirements of this use case. Whereas the
requirement 1 is purely structural, the requirements 2 and 3 are
relative to the execution of the SoS: the first one is written using
strictly OCL, the second one shows the cumulative operators we
introduced and the third one is defined with a behavioral pattern.
The presented requirements are contracts without assumption or, more precisely,
they are contracts with an assumption that is implicitly "true".

\newcommand{\inblack}[1]{{\bf #1}}
\newcommand{\inred}[1]{{\bf \color{red}{#1}}}
\newcommand{\inblue}[1]{{\bf \color{blue}{#1}}}
\newcommand{\sep}{$\inblue{|}$}

\begin{table}[h]
  \centering
  \hspace*{-3mm}
  \begin{tabular}{|l|}
    \hline

    \emph{"Any district cannot have more than 1 fire station, except if all districts have at least 1"}\\
    \inred{SoS.itsDistricts}\inblue{$\rightarrow$exists(}district \sep{} district.\inred{containedFireStations}\inblue{$\rightarrow$size() $>$ 1)} \inblue{implies}\\
    \hspace*{1cm} \inred{SoS.itsDistricts}\inblue{$\rightarrow$forAll(}district \sep{} district.\inred{containedFireStations}\inblue{$\rightarrow$size() $\ge$ 1)}\\
    \hline
    \emph{"The mean city area under fire shall be less than 0.01\%"}\\

    \inblack{mean}(\inred{SoS.itsDistricts.fireArea}\inblue{$\rightarrow$sum()}) $\leq$ 0.0001\\
    \hline

    \emph{"The fire fighting cars hosted by a fire station shall be used all simultaneously at least once}\\
    \hspace*{3mm}\emph{in 6 months"}\\
    \inred{SoS.itsFireStations}\inblue{$\rightarrow$forAll(}fireStation \sep{}\\
    \hspace*{1cm} \inblack{Whenever [}fireStation.\inred{hostedFireFightingCars}\inblue{$\rightarrow$exists(}ffCar \sep{} ffCar.\inred{isAtFireStation}\inblue{)}\inblack{] occurs,}\\
    \hspace*{2.8cm} {\bf [}fireStation.\inred{hostedFireFightingCars}\inblue{$\rightarrow$forall(}ffCar \sep{} ffCar.\inred{isAtFireStation} \inblue{= false)}\inblack{]}\\
    \hspace*{12cm} \inblack{occurs within [6 months]}\inblue{)}\\
    \hline
  \end{tabular}

  \caption{Examples of Requirements formulated in the CAE incubator}
  \label{tab:requirements_CAEincubator}
\end{table}

The proposed language ,composed by 11 SPEEDS patterns, is sufficient expressive to formalize the behavior from 15 requirements identified in CAE incubator. This list of patterns can be easily extended for the future needs, but the experiments conducted in SPEEDS and {\sc Danse} show it covered all the requirements to be expressed.


\section{Translating Contracts into Bounded-LTL Formulas}
\label{sec:translation}

\subsection*{Bounded Linear Temporal Logic}

As said previously, the Bounded Linear Temporal Logic
(B-LTL) is an extension of the Linear Temporal Logic
(LTL)~\cite{CGP99} in which each temporal operator is bound by a
temporal constant. This Logic is such expressive that it covers
precisely a large set of properties. It is particularly adapted to
Statistical Model Checking (SMC)~\cite{Younes05, KVA05}. The SMC principle is to monitor some
simulations in order to check a B-LTL property and use the results
from the statistics area (sequential hypothesis testing or Monte Carlo
simulation) in order to decide whether the system satisfies the B-LTL
property or not with some degree of confidence. Since the conducted
simulations are finite, the infinite path semantics of LTL has no
sense, whereas checking B-LTL formulas does.

The formulas are built using the standard logic connectors $\land$,
$\lor$, $\implies$, $\neg$ and the common temporal modalities $G$,
$F$, $X$, $U$ over some atomic propositions. Each temporal modality is
indiced by a bound defining the length of the run on which the
formula must hold. The validation of a B-LTL formula against an
execution path has a meaning only if the length of this path is
enough to reach all bounds constituting the formula.

The atomic propositions used in the B-LTL formulas are build using
some state predicates or run predicates. These predicates only require
to be decidable for a given input, e.g. a state or a run section, and
we assume this decision to be performed by an external
procedure. Considering $\pi = s_0s_1\dots s_n$ a finite run of a
transition system and $\Phi$ a B-LTL property, $\pi \models \Phi$
means that the run $\pi$ satisfies the property $\Phi$. The suffix
$s_is_{i+1}\dots s_n$ of $\pi$ is noted $\pi^i$. Assuming $k > 0$, a
run $\pi = s_0s_1\dots s_n$, a state predicate $P$ and a run predicate
$Q$, the satisfiability of the B-LTL formulas $\Phi$, $\Phi_1$ and
$\Phi_2$ is defined in Table~\ref{tab:semantics}.

\begin{table}[h!]
  \[
  \hspace*{-4mm}
  \begin{array}{|lclp{1.2cm}|}
  \hline
  \pi \models F_{\le k} \Phi &\equiv& \exists i,\ t_0 \le t_i \le t_0 + k \quad\textrm{and}\quad \pi^i \models \Phi &\\

  \pi \models G_{\le k} \Phi &\equiv& \forall i,\ t_0 \le t_i \le t_0 + k \quad\textrm{and}\quad \pi^i \models \Phi&\\

  \pi \models X_{\le k} \Phi &\equiv& \forall i,\ i = max\{j\ |\ t_0 \le t_j \le t_0 + k\} \quad\textrm{and}\quad \pi^i \models \Phi &\\

  \pi \models \Phi_1 \ U_{\le k} \Phi_2 &\equiv& \exists i,\ t_0 \le t_i \le t_0 + k \quad\textrm{and}\quad\pi^i \models \Phi_2 \quad\textrm{and}\quad \forall j, 0 \le j \le j,\ \pi^j \models \Phi_1 &\\

  \pi \models \Phi_1 \ W_{\le k} \Phi_2 &\equiv& \pi \models  (\Phi_1 \ U_{\le k} \Phi_2) \lor  G_{\le k} \Phi_2&\\

  \pi \models \Phi_1 \implies \Phi_2 &\equiv& \pi \models \neg\Phi_1 \lor \Phi_2&\\
  \pi \models \Phi_1 \lor\Phi_2 &\equiv& \pi \models \Phi_1 \quad\textrm{or}\quad \pi \models \Phi_2&\\
  \pi \models \Phi_1 \land \Phi_2 &\equiv& \pi \models \Phi_1 \quad\textrm{and}\quad \pi \models \Phi_2&\\
  \pi \models \neg\Phi   &\equiv& \pi \not\models \Phi&\\
  \pi \models P &\equiv& P(s) \textrm{ holds} & \hspace*{-4cm}\textrm{checked by an external procedure}\\
  \pi \models Q &\equiv& Q(\pi) \textrm{ holds}&\\
  \pi \models true&&&\\
  \pi \not\models false&&&\\
  \hline
\end{array}
\]
  \caption{Semantics of B-LTL}
  \label{tab:semantics}
\end{table}

\begin{example}[Example of B-LTL formula]
  Let us consider the formula $G_{\le 5} (A \implies X_{\le 1}F_{\le 2} B)$
  where $A$ and $B$ are state propositions and an execution path $\pi$
  such that $A$ and $B$ hold as illustrated below:

\end{example}

\subsection*{Overview of the translation procedure}

As illustrated in the third example of requirements of
Table~\ref{tab:requirements_CAEincubator} the language is layered as
some behavioral properties defined using the patterns combined with
some state properties written in OCL These behavioral properties can
themselves be wrapped into an OCL collection expression to quantify
the behavioral properties over some constituents of the SoS.
The translation of a contract will be made by translating from its assumption
and its promise only the OCL quantification and the pattern
layers.  The translated property will be checked
against some simulations. The state properties expressed in OCL have
to be checked against some states and for them, no treatment is done
during the translation. The state properties are kept in the
translated formula and there will be dynamically checked. We assume
that the satisfiability of the state properties is solved by an
external procedure based on an existing OCL-checker~\cite{OCLChecker}.

\begin{proposition}
  Let us consider a contract $(A, P)$ of a given SoS and assume any
  simulation is bounded by $k$ a maximum time of execution.  If there
  exist two B-LTL formulas $A'$ and $P'$ such that $A'$ (or $P$) and
  $A$ ($P'$ resp.) are equivalent for any $k$-bounded simulations,
  then the B-LTL formula $A' \implies P'$ is equivalent to the
  contract $(A, P)$ for any $k$-bounded simulations.
\end{proposition}
The proof is a trivial consequence of Definition~\ref{def:contract}
written using B-LTL.  Moreover, extending the translation to a
stochastic contract is natural. The pair $(A, P)$ of any stochastic
contract is similarly treated.

 \subsection*{OCL quantification translation}
 OCL expressions occur at two levels within a pattern: as atomic
 propositions to define a state condition and as quantifications. The
 first case will be directly treated by an external OCL-checker against a
 state of the SoS or translated into a more generic semantics provided
 by the SMC-checker. But some atomic propositions can also contain
 some quantification about component collections and in this case they
 can also be processed as explained below.  The second case is the
 most interesting case. The B-LTL logic has no quantification support;
 it could be extended but this needs to rewrite the B-LTL
 checker. Moreover, adding quantifications to the logic increases
 significantly the complexity of the satisfiablity decision.

 Moreover, the instances of each component type are statically
 specified in the Internal Block Diagram (IBD) by the SoS
 architect. In the CAE incubator, the IDB is named {\tt
   idbFireEmergency} and gives the list of all system constituents
 instantiated in the SoS: 10 districts, 1 fire headquarter, 3 fire
 stations and 7 fire fighting cars shared by the fire stations,
 etc. Since the number of constituents is known and finite in the SoS,
 any universal quantification (or an existential quantification) over
 a collection can be interpreted as a conjunction (a disjunction
 resp.). Using the CAE incubator and assuming a valid property $\phi$
 for the fire stations, the property ${\tt
   SoS.itsFireStations\rightarrow forAll}({\tt x}|\ \phi({\tt x}))$ is
 equivalent to $\phi({\tt fireStation_1}) \land \phi({\tt
   fireStation_2}) \land \phi({\tt fireStation_3})$.


In cases where $\phi$ contains also a quantification, $\phi$ must also be unfold. The generalization of the process
is recursively defined as:
\[
\newcommand{\niceConnector}[1]{
  \begin{minipage}[h]{.2\linewidth}
    \vspace*{-4mm}
    \[#1_{\tt x \in coll} unfold\big(\phi({\tt x})\big)\vspace*{1mm}\]
  \end{minipage}}
\left\{\begin{array}{lcl}
  unfold\big({\tt coll\rightarrow forAll(x | \phi(x)}\big)&= &\niceConnector{\bigwedge}\\
  unfold\big({\tt coll\rightarrow exists(x | \phi(x)}\big)&= &\niceConnector{\bigvee}\\
  unfold\big({\tt expr}) & = & {\tt expr}, \quad \textrm{otherwise}\\
\end{array}\right.
\]
where {\tt coll} is an OCL collection and {\tt expr} any other valid expression of the contract language.

\subsection*{Pattern translation}
For the purpose of translation to B-LTL, we assume that the constant
$k$, is an additional parameter given by the user: it corresponds to
the execution time for which we expect the property hold.  Whenever an
unbound pattern to write a property like "Always \dots" is meaning
full for the specification, statistical model checking still checks
the B-LTL property against a simulation that is finite: $k$ is used to
replace the implicit unbound value of the property.  Moreover, to be
successfully translated, the pattern must be consistent: in particular
the intervals must have correct bounds and intervals must be all
visited before the end of the simulation (and so the constant $k$) is
reached.

\begin{proposition}
  Assuming $\Psi$, $\Psi_1$ and $\Psi_2$ denote some state propositions
  nested in a pattern $P$, e.g. OCL propositions, and a constant $k > 0$ then, for any run of length $k$,
  there exists a B-LTL formula equivalent to $P$. Table~\ref{tab:pattern} summarizes the valid B-LTL translations.
\end{proposition}

\begin{table}[h]
  \centering
  \newcounter{patterncounter}
  \newcommand{\idPattern}{\stepcounter{patterncounter}\arabic{patterncounter}}
  \newcommand{\jump}{\hspace*{3mm}}
  \begin{tabular}{|c|p{5.5cm}|l|r|}
    \hline
    \multicolumn{2}{|c} {} &
    \multicolumn{1}{|c|} {} &
    \multicolumn{1}{c|}{Consistency}\\
    \multicolumn{2}{|c}  {Pattern} &
    \multicolumn{1}{|c|} {B-LTL translation} &
    \multicolumn{1}{c|}  {Condition}
     \\\hline
     \multicolumn{4}{c}  {\jump {\sc Basic B-LTL patterns with absolute intervals}}
     \\\hline
     \idPattern&\kw{always} $\Psi$ & $G_{\le k} \Psi$ & -\hspace*{9mm}
      \\\hline
     \idPattern&\kw{whenever} $\Psi_1$ \kw{occurs} $\Psi_2$ \kw{holds} & $G_{\le k}(\Psi_1 \implies \Psi_2)$ & -\hspace*{9mm}
      \\\hline
     \idPattern&$\Psi_1$ \kw{implies} $\Psi_2$ \kw{during following} $[a,b]$ &$X_{\le a} G_{\le a - b}(\Psi_1 \implies \Psi_2)$ &$a \le b$
      \\\hline
     \idPattern&$\Psi_1$ \kw{during} $[a, b]$ \kw{raises} $\Psi_2$&
     $(X_{\le a}G_{\le a - b} \Psi_1) \implies X_{\le b} \Psi_2$ & $a \le b \le k$
      \\\hline
     \idPattern&$\Psi$ \kw{during} $[a, b]$ \kw{implies} $\Psi_1$ \kw{during} $[a, c]$ \kw{then} $\Psi_2$ \kw{during} $[c,b]$
      &$X_{\le a}G_{\le b-a}\big(X_{\le a}G_{\le c-a}(\Psi_1) \land X_{\le c}G_{\le b-c}(\Psi_2)\big)$ & $a \le c \le b$
      \\\hline

      \multicolumn{4}{c} {\jump {\sc Extended B-LTL patterns with absolute intervals}}
     \\\hline

     \idPattern&$\Psi_1$ \kw{occurs} $n$ \kw{times during} $[a, b]$ \kw{raises} $\Psi_2$
      &$occ(\Psi_1, a, b) \ge n \implies X_{\le b} F_{\le k - b} \Psi_2$& $a\le b\le k$
      \\\hline

     \idPattern&$\Psi$ \kw{occurs at most} $n$ \kw{times during}~$[a,b]$ & $occ(\Psi, a, b) \le n$ & $a \le b$
      \\\hline
      \multicolumn{4}{c} {\jump {\sc Basic B-LTL patterns with sliding intervals}}
     \\\hline

     \idPattern&\kw{whenever} $\Psi_1$ \kw{occurs} $\Psi_2$ \kw{holds during following} $[a,b]$
      &$G_{\le k - b}(\Psi_1 \implies X_{\le a}G_{\le b - a} \Psi_2)$ & $a \le b \le k$
     \\\hline
    \idPattern&\kw{whenever} $\Psi$ \kw{occurs} $\Psi_1$ \kw{implies} $\Psi_2$ \kw{during following} $[a, b]$
      &$G_{\le k - b}(\Psi \implies X_{\le a}G_{\le b - a} (\Psi_1 \implies \Psi_2))$ & $a \le b \le k$
     \\\hline
    \idPattern&\kw{whenever} $\Psi_1$ \kw{occurs} $\Psi_2$ \kw{does not occur during following} $[a, b]$
      &$G_{\le k - b}(\Psi_1 \implies X_{\le a}G_{\le b - a} \neg \Psi_2)$ & $a \le b \le k$
     \\\hline
    \idPattern&\kw{whenever} $\Psi_1$ \kw{occurs} $\Psi_2$ \kw{occurs within} $[a,b]$
      &$G_{\le k - b}(\Psi_1 \implies X_{\le a}F_{\le b - a} \Psi_2)$ & $a \le b \le k$
     \\\hline
   \end{tabular}
   \caption{\small Pattern mapping}

   \label{tab:pattern}
 \end{table}

 \paragraph{Unbound time patterns} The patterns 1 and 2 require that
 the expressed properties must hold while the system is running,
 e.g. they have a meaning for infinite execution paths too. But, the
 verification will be done against simulation path that are
 necessarily finite, for practical reasons (termination). Thus, the
 infinite bound is replaced by the user constant $k$ provided for the
 verification.
 \paragraph{Extended B-LTL patterns} The patterns 6 and 7 require to
 count the number of occurrences in $[a, b]$. Counting is not
 possible by strictly using B-LTL. We assume that there exist a
 dedicated procedure $occ(\Psi,a,b)$ that counts the number of times
 where $\Psi$ is satisfied and compare it to the value $n$. We use similarly
 some external treatment to evaluate the operators $sum()$, $mean()$, \dots
 that compute a accumulated value of an expression during a time interval.
 \paragraph{Sliding intervals} the interval $[a, b]$ to consider is
 located after time $t$ at which the first part of the pattern "{\bf Whenever} $\Psi$ {\bf occurs}" is
 satisfied: this sliding interval in pattern 8 is encoded as the property $\Psi_2$ holds during the duration $b - a$ after $a$ units of time after we observe $\Psi_1$ is true.

\subsection*{Illustration of the full translation}
We illustrate the translation for the third requirement in Table~\ref{tab:requirements_CAEincubator}.

\hspace*{-7mm}
\fbox{
\begin{tabular}{l}
  \inred{SoS.itsFireStations}\inblue{$\rightarrow$forAll(}fireStation $|$\\
    \hspace*{1cm} \inblack{Whenever [}fireStation.\inred{hostedFireFightingCars}\inblue{$\rightarrow$exists(\inred{isAtFireStation})}\inblack{] occurs,}\\
    \hspace*{2.8cm} {\bf [}fireStation.\inred{hostedFireFightingCars}\inblue{$\rightarrow$forall(\inred{isAtFireStation} = false)}\inblack{]}\\
    \hspace*{11cm} \inblack{occurs within [6 months]}\inblue{)}\\
\end{tabular}
}

Assuming that we have the time bound $k \ge 6months$ the pattern is translated to the B-LTL formula following the rule 12 in Table~\ref{tab:pattern}:

\[
\newcommand{\theProp}[1]{G_{\le k - 6months}(\Psi_1(#1) \implies X_{\le 0}F_{\le 6months} \Psi_2(#1))}
\phi = \left\{
  \begin{array}{c}
    \theProp{{\tt fireStation_1}}\\
    \bigwedge\\
    \theProp{{\tt fireStation_2}}\\
    \bigwedge\\
    \theProp{{\tt fireStation_3}}\\
  \end{array}
\right.
\]

where $\Psi_1({\tt fireStation_i})$ and $\Psi_2({\tt fireStation_i})$ correspond to the OCL expressions in brackets, e.g. \makebox[1mm]{ } ${\tt fireStation.hostedFireFightingCars\rightarrow exists(isAtFireStation)}$\makebox[0.5mm]{ } and ${\tt fireStation.}$\\ ${\tt hostedFireFightingCars\rightarrow forall(isAtFireStation = false)}$. We notice that the modality $X_{\le 0}$ could be cleaned in $\phi$, but we leave it for the sake of clarity.

As for the OCL quantification at the root of the requirement, we unfold the OCL quantifications that occur in $\Psi_1$ and $\Psi_2$.  The next table gives the result of this unfolding in $\Psi_1({\tt fireStation_i})$ and $\Psi_2({\tt fireStation_i})$ for each ${\tt fireStation}$. Finally, replacing all occurences of $\Psi_1({\tt fireStation_i})$ and $\Psi_2({\tt fireStation_i})$ in $\phi$ gives the complete translation in B-LTL.

\begin{table}[h]
  \centering
{\tt \small
\hspace*{-3mm}\begin{tabular}{|l|l|l|}
  \hline
  \multicolumn{1}{|c} {Component} &
  \multicolumn{1}{|c|} {$\Psi_1$} &
  \multicolumn{1}{c|} {$\Psi_2$} \\
  \hline
  fireStation1 & fireFightingCar1.isAtFireStation $\lor$ & $\neg$ fireFightingCar1.isAtFireStation $\land$ \\
  &fireFightingCar2.isAtFireStation $\lor$ &$\neg$ fireFightingCar2.isAtFireStation $\land$ \\
  &fireFightingCar3.isAtFireStation & $\neg$ fireFightingCar3.isAtFireStation \\
  \hline
  fireStation2 & fireFightingCar4.isAtFireStation $\lor$ & $\neg$ fireFightingCar4.isAtFireStation $\land$ \\
  & fireFightingCar5.isAtFireStation & $\neg$ fireFightingCar5.isAtFireStation \\
  \hline
  fireStation3 & fireFightingCar6.isAtFireStation $\lor$ & $\neg$ fireFightingCar6.isAtFireStation $\land$ \\
  & fireFightingCar7.isAtFireStation & $\neg$ fireFightingCar7.isAtFireStation \\
  \hline
\end{tabular}
}
\end{table}

\section{Statistical Model Checking of SoS Contracts}
\label{sec:stat-model-check}

\begin{figure}[h]
  \centering
  \includegraphics[width=9.5cm]{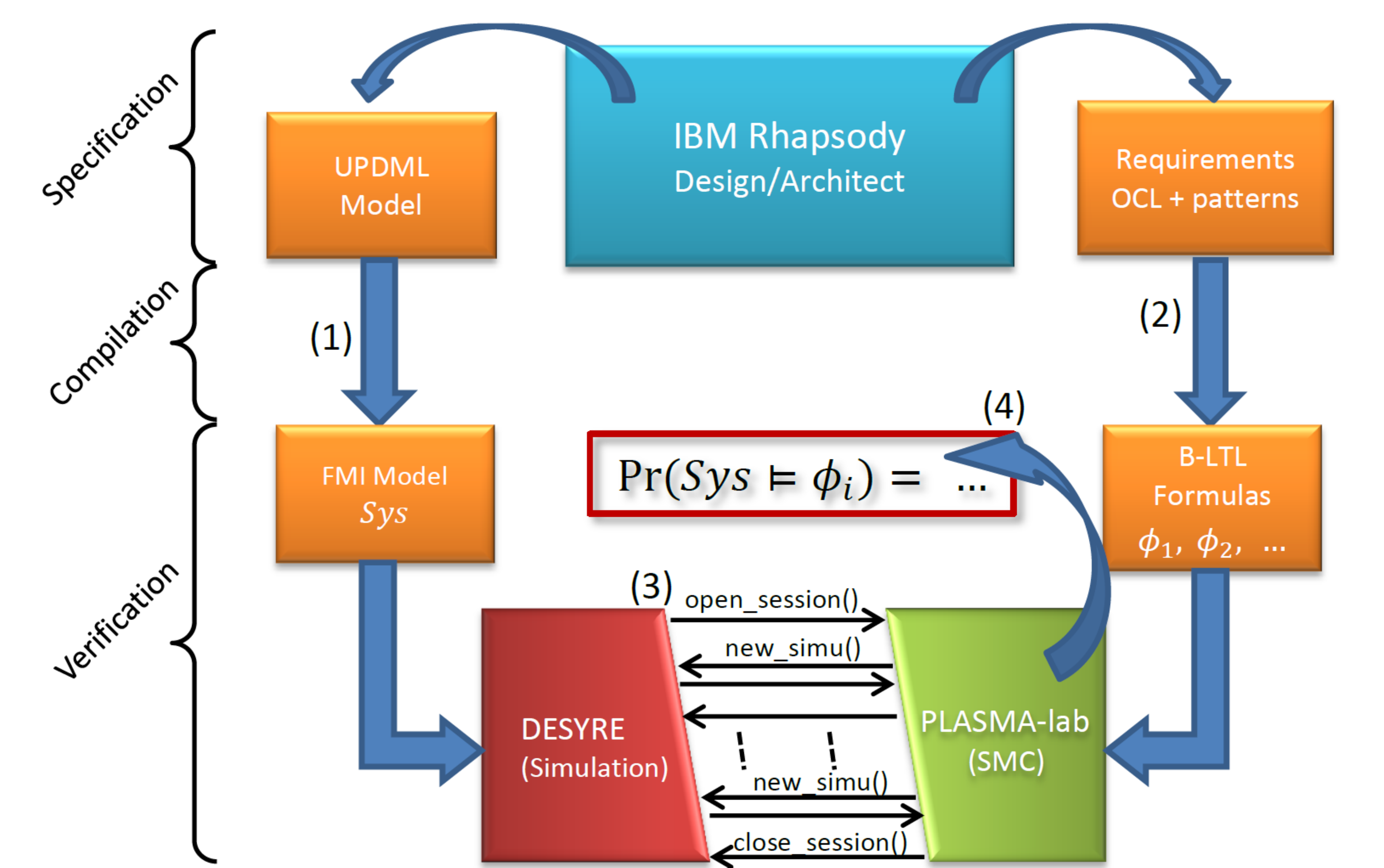}
  \caption{The SMC process in {\sc Danse}}
  \label{fig:SMC-process}
\end{figure}

The interest of SMC~\cite{Younes05, KVA05} is to propose an
alternative to the approach of the classical model
checking~\cite{BK08,CGP99}.  By using results from the statistic area
(including sequential hypothesis testing or Monte Carlo simulation) in
order to decide whether the system satisfies the property or not with
some degree of confidence, SMC avoids an exhaustive exploration of the
state-space of the model that generally does not scale up.  It has
already successfully experimented in biology
area~\cite{CFLHJL08,JCLLPZ09,PRISM}, software engineering~\cite{CLD10}
as well as industrial area\cite{BBBCDL10} More recently, in {\sc
  Danse}~\cite{project-DANSE}, we adapt the SMC techniques to treat
large heterogeneous systems like Systems of Systems. Among them, one
finds systems integrating multiple heterogeneous distributed
applications communicating over a shared network. We proposed to
extend UPDM specification - the SoS specification - with some
requirements that the SoS must satisfy. These requirements, are
specified with the contract language we specially designed for the
SoS's. These goals are viewed as behavioral objectives that support
the SoS architect in assessing different strategies and finding the
best ones. As shown in Figure~\ref{fig:SMC-process}, these contracts
are compiled into B-LTL formulas that are verified against the SoS
(whose constituent systems are compiled into FMI executables) using
the Statistical Model Checker Plasma-Lab~\cite{Plasma-Lab} combined
with the efficient simulation engine DESYRE developed by
Ales~\cite{Ales}. The SMC tool-chain gives an estimation of the
satisfiability of the contract by the SoS. The different results help
the SoS architect to make good decisions about how to optimize the SoS
strategies.

The main algorithm we used in {\sc Danse} is the Monte Carlo algorithm.
This algorithm estimates the probability that a system $Sys$ satisfies
a B-LTL property $P$ by checking $P$ against a set of $N$ random
executions of $SyS$. The estimation $\hat{p}$ is given by
\[\hat{p} = \dfrac{\sum_1^N f(ex_i)}{N} \quad \textrm{where } f(ex_i) = 1 \textrm{ if } ex_i\models P,\ 0 \textrm{ otherwise} \]

Using the formal semantics of B-LTL, each execution trace is monitored in order to check if $P$
is satisfied or not. The accuracy of the estimation increases with a bigger number of monitored simulations.

Plasma-Lab\cite{Plasma-Lab} implements a set of tools from the statistical area to perform the SMC. It provides some engines for simulating biologic models, models written in the Prism language~\cite{PRISM}, but it has also the capabilities to drive an external engine to perform the simulations like MathLab, SciLab, or DESYRE.


\section{Illustration using the CAE incubator}
\label{sec:illustration}

In the frame of the {\sc Danse} project, the Concept Alignment Example
(CAE) is a fictive SoS example inspired by real-world Emergency
Response data to a city fire. It has been built as a playground to
demonstrate new methods and models for the analysis and visualization
of SoS designs. All structural modeling has been performed using UPDM
views, and behaviors have been added on a subset of the constituents
that we called "CAE incubator", using simple SysML constructs
(modeled in state machines) extended by a few stereotypes (e.g. for
storing stochastic information).

Behavioral modeling in the CAE incubator is focused on following
constituent systems: Fire HQ, Fire Station, Fire Fighting Car and
District. The city districts have been added as constituent systems
because they play an important role in the SoS: their behavior
describes how the fires arise, expand and spread to neighbor
districts. In the frame of the CAE, all behaviors are abstracted in
state machines using IBM Rhapsody, but it would be possible to use any
other language and tool as long as it is compliant with the FMI export
format.

The following figure shows the overall architecture of the CAE
incubator as well as the behavior of one of the constituent types: a
fire fighting car.

\begin{figure}
\centering
\includegraphics[width=0.9\linewidth]{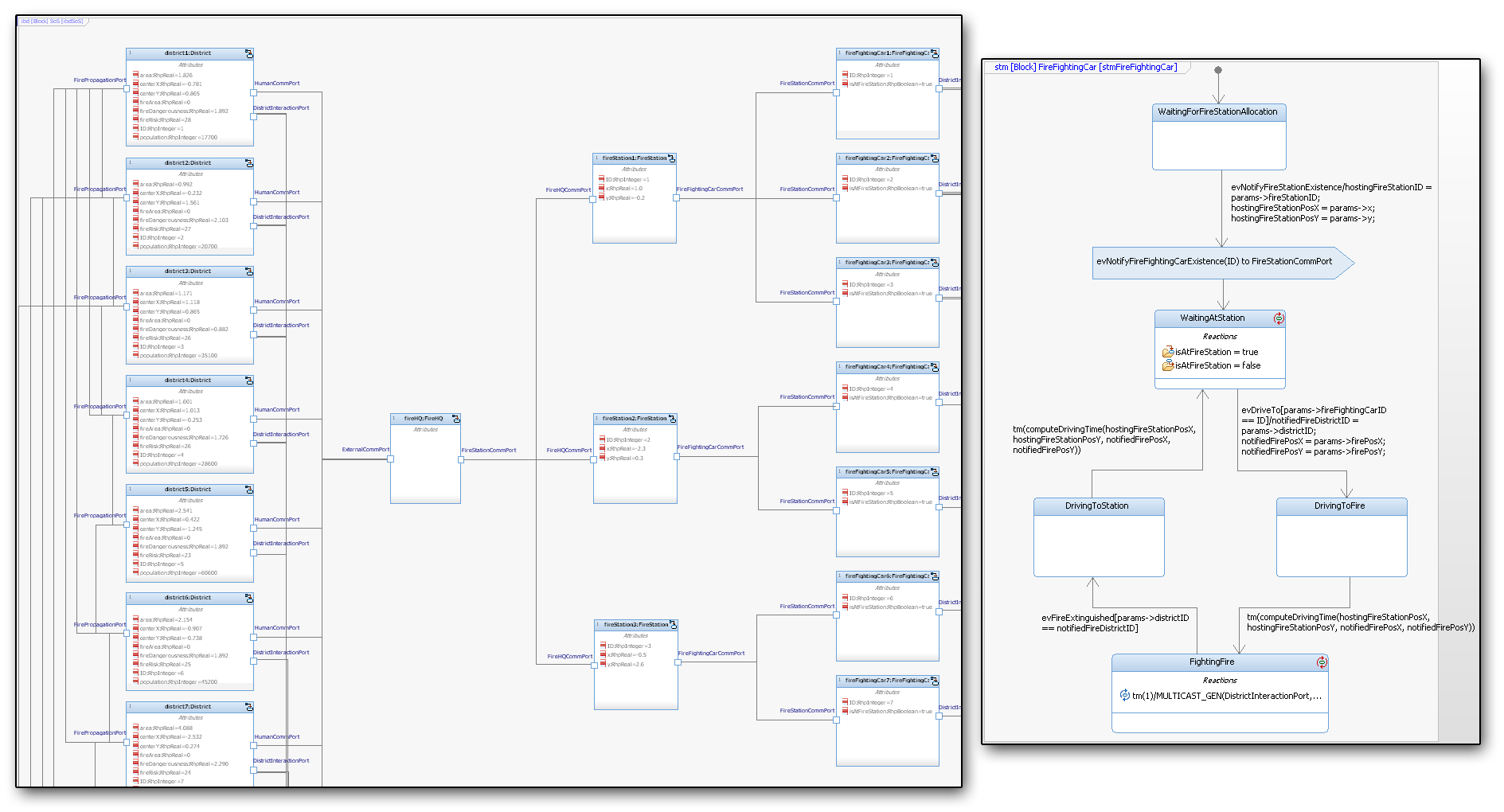}
\caption{CAE incubator - architecture example and behavior of a fireman}
\end{figure}

We attached to the CAE incubator the following requirement, written
accordingly to our proposed formalism:

\begin{table}[h]
  \centering
  \begin{tabular}{|l|}
    \hline
    \emph{"The mean city area under fire shall be less than 0.01\%"}\\
    mean(\inred{SoS.itsDistricts.fireArea}\inblue{$\rightarrow$sum()}) $\leq$ 0.01 \%\\
    \hline
  \end{tabular}
\end{table}

As described in the paper, we were able to translate this requirement to the low-level B-LTL specification for the statistical model checker Plasma-Lab and use it in conjunction with the simulation platform DESYRE to assess the probability that this goal is met in the specified time range (the simulation time for each run was 4 months). By choosing the Monte Carlo option, Plasma-Lab was able to give us the following estimation as a result on a given number of runs:\\
\begin{center}
  \fbox{$Prob(\textrm{mean city area under fire} \leq 0.01 \%) \approx 92.3 \% $}\\
\end{center}

In addition to the computation of the estimated probability that this goal is met on a given number of runs, Plasma-Lab can also compute how many runs are necessary to prove that a given probability threshold is passed by choosing the Chernov option.

\section*{Conclusion}
This papers presents the results of the very first contract-based
language for UPDM/SysML model of SoS we developed in the {\sc Danse}
project. The SoS model used in the project remains rather simple, but
powerful enough to capture behaviors and requirements of a CAE case
study developed in collaboration with EADS. Also, we are the first to
study the relation between a modeling language used in industry (UPDM)
and a verification approach developed by academic.

As a future work, we plan to offer more dynamicity, which we will do
by exploiting and extending the work done on adaptative
systems~\cite{ZC06,Ghezzi11,ChengLGIMABBBCSDFGGGKKKLMMMPSTTWW09,FHNPSV11}. This
will also requires to adapt the UPDM framework.

Another interesting future work will be to add more quantitative information directly in the patterns assumption and guarantee. This will permit us to reason on complex problematic such as energy consumption.

All these future extensions will be discussed and
designed jointly with the business units as the {\sc Danse} partners.



\nocite{*}
\bibliographystyle{eptcs}
\bibliography{biblio}

\newpage
\appendix

\section{Grammar}

\setlength{\grammarparsep}{20pt} 
\setlength{\grammarindent}{7em} 

\newcommand{\Nat}{\mathbb{N}}
\newcommand{\F}{\mathbb{F}}
\begin{grammar}

<contract> ::= {<viewpoint-id>}+ `contract' <identifier>
\{`Assumption:' <property>\} ?
`Goal:' <property>
`Confidence:' <threshold>

<viewpoint-id> ::= `dynamicity' | `behavior' | `structure' | `safety' | `liveness' | \dots

<threshold> ::= $\mathit{F}loat$`\%' | <probability>

<probability> ::= $x,\ x \in (0; 1]$

<property> ::= <OCL-coll> `->forAll('<variable> `|' <pattern>`)'
\alt <OCL-coll> `->exists('<variable> `|' <pattern>`)'
\alt <OCL-prop>
\alt <pattern>

<pattern> ::=
     `whenever' `['<prop>`]' `occurs' `['<prop>`]' `holds' `during' `following' `['<int>`]'
\alt `whenever' `['<prop>`]' `occurs' `['<prop>`]' `implies' `['<prop>`]' `during' `following' `['<int>`]'
\alt `whenever' `['<prop>`]' `occurs' `['<prop>`]' `does' `not' `occur' `during' `following' `['<int>`]'
\alt `whenever' `['<prop>`]' `occurs' `['<prop>`]' `occurs' `within' `['<int>`]'
\alt `['<prop>`]' `during' `['<int>`]' raises `['<prop>`]'
\alt `['<prop>`]' `occurs' `['$\Nat$`]' times during `['<int>`]' `raises' `['<prop>`]'
\alt `['<prop>`]' `occurs' `at' `most' `['$\Nat$`]' `times' `during' `['<int>`]'
\alt `['<prop>`]' `during' `['<int>`]' `implies' `['<prop>`]' `during' `['<int>`]' `then' `['<prop>`]' `during' `['<int>`]'

<prop> ::= <OCL-prop> | <arith-rel>

<arith-rel> ::= <expr> ( `<' | `<=' | `=' | `>=' | `>' ) <expr>

<arith-expr> ::= <expr> <operator> <expr> | `('<expr>`)'
\alt <OCL-expr>
\alt `mean(' <OCL-expr> `)' | `sum(' <OCL-expr> `)'
\alt `prod(' <OCL-expr> `)' |  `at(' <OCL-expr>`,' <time> `)'

<operator> ::= `+' | `-' | `*' | `/'

<int> ::= \{`[' | `('\} <time> \{`-' <time>\}? \{`]' | `)'\}

<time> ::= $\Nat$ <time-unit> | $+\infty$

<time-unit> ::= `ms' | `s' | `min' | `hour' | `day' | \dots
\end{grammar}

\newcommand{\nonTerm}[1]{$\langle${\it #1}$\rangle$}

The non-terminal \nonTerm{time-unit} can be any multiple of the application basic time unit ( i.e. day, hour, min, sec, ms, ...).
The latest revision of the OCL specification can be found at~\cite{OCL} and more particularly the grammar of the language. We just give an overview of the relevant subset used in this language:
\nonTerm{OCL-proposition} stands for the simple Boolean expressions over collections or primitive types (int, real,
boolean, \dots) of OCL. We also identified \nonTerm{OCL-expr}, the OCL subset of non-Boolean expression, e.g.
Component Collections (without treatments, e.g. the functions {\tt map(\dots), iter(\dots)} ), numerical values, model-related values, \dots Some relevant details about OCL collections are in the chapters 7.7 (Collection operations) and 11.6 (Collection-related types) of the OCL specification~\cite{OCL}.


\section{Patterns}
\label{sec:patterns}

We give the list of all the SPPEDS patterns we reuse and we give give their semantics based on the statifiability given in Section~\ref{sec:contr-lang-updmlsysm}.

\begin{enumerate}[label=\alph*.]
\item \kw{whenever} $\Psi_1$ \kw{occurs} $\Psi_2$ \kw{holds during following} $[a,b]$
  \begin{center}
    \includegraphics[width=.6\linewidth]{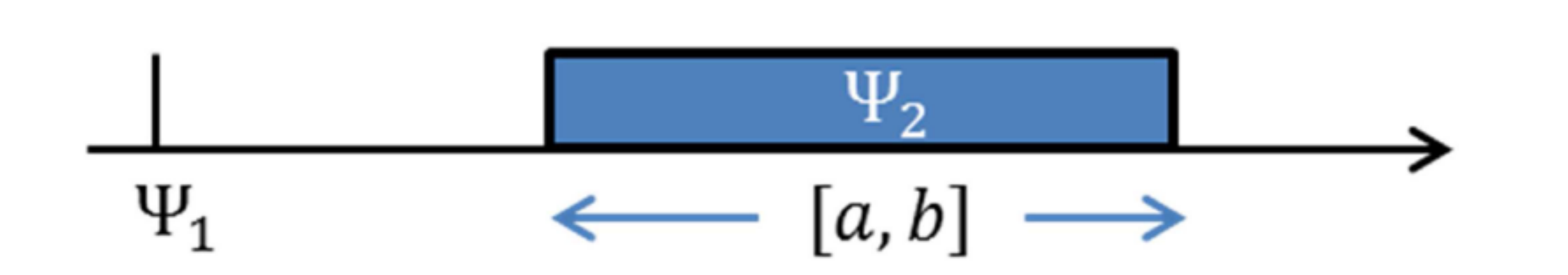}
  \end{center}
The interval $[a, b]$ is located relatively after the satisfaction of $\Psi_1$ .
The interval, in which $\Psi_2$ must be satisfied, starts $a$ units of time after
the observed occurrence of $\Psi_1$.

\item $\Psi_1$ \kw{implies} $\Psi_2$ \kw{holds forever}
  \begin{center}
    \includegraphics[width=.6\linewidth]{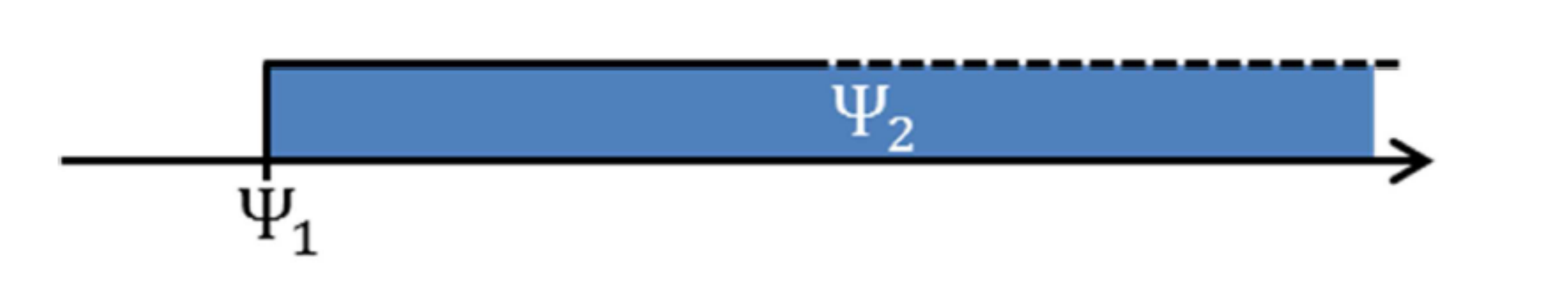}
  \end{center}
From the very moment when $\Psi_1$ is satisfied $\Psi_2$ must hold during all the rest of the execution path.

\item \kw{always} $\Psi$
  \begin{center}
    \includegraphics[width=.6\linewidth]{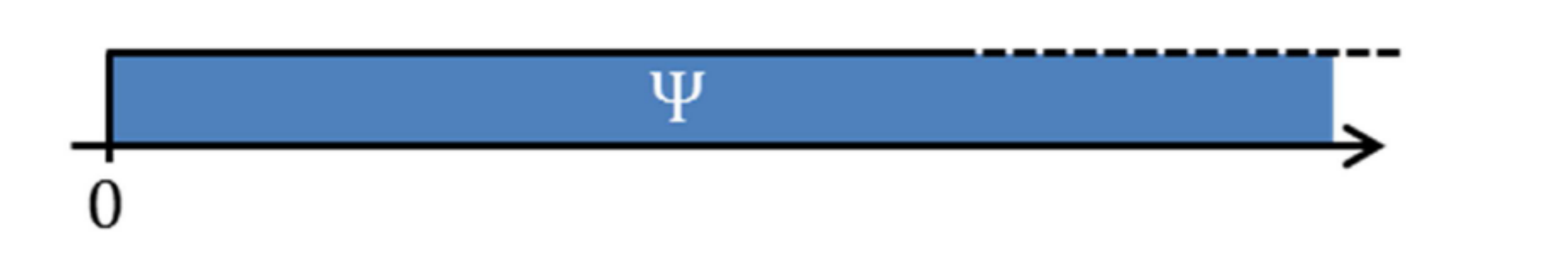}
  \end{center}
$\Psi$ must hold during all the execution path.

\item \kw{whenever} $\Psi_1$ \kw{occurs} $\Psi_2$ \kw{holds}
    \begin{center}
    \includegraphics[width=.6\linewidth]{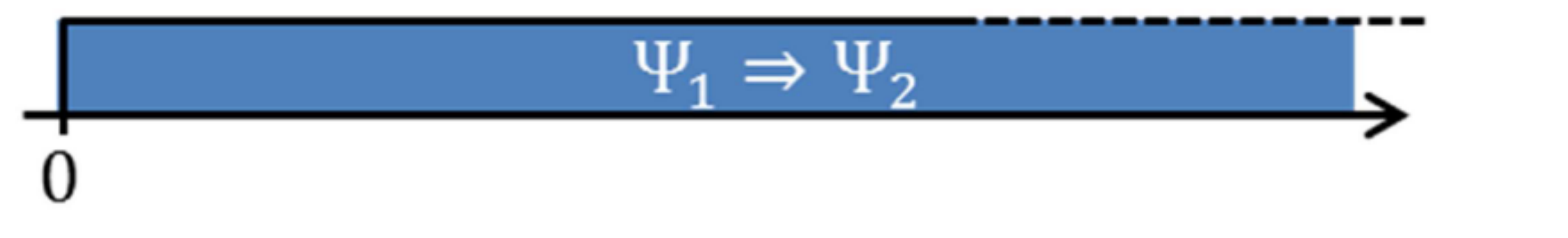}
  \end{center}

\item \kw{whenever} $\Psi$ \kw{occurs} $\Psi_1$ \kw{implies} $\Psi_2$ \kw{during following} $[a,b]$
  \begin{center}
    \includegraphics[width=.6\linewidth]{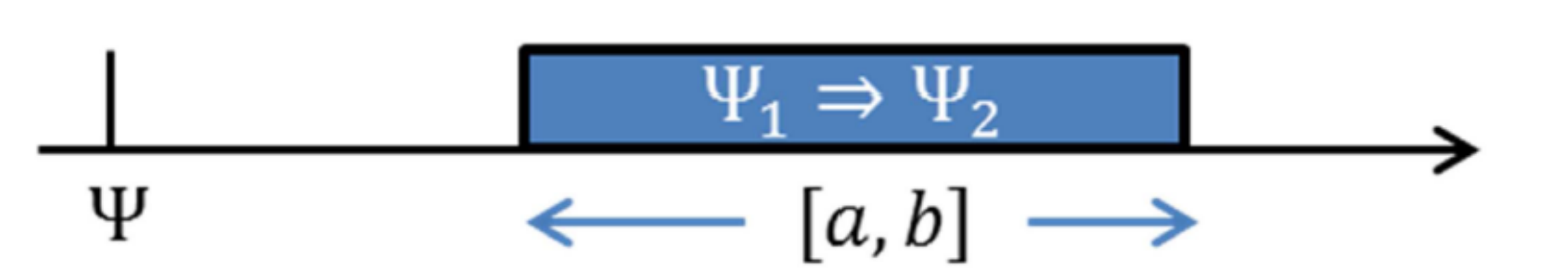}
  \end{center}
  As for the previous pattern, the interval $[a, b]$ is relative. At
  each time value between $a$ and $b$.  where $\Psi_1$ holds, $\Psi_2$
  must also hold. Replacing $\Psi$ is replaced by $true$ allows to create a new simpler pattern:\\
  $\Psi_1$ \kw{implies} $\Psi_2$ \kw{during following} $[a, b]$\\

\item \kw{whenever} $\Psi_1$ \kw{occurs} $\Psi_2$ \kw{does not occur during following} $[a,b]$
    \begin{center}
    \includegraphics[width=.6\linewidth]{figs/pattern_f}
  \end{center}
  This pattern specifies that $\Psi_2$ is never satisfied during the
  relative interval $[a, b]$, i.e. $\neg\Psi_2$ holds during $[a, b]$.

\item \kw{whenever} $\Psi_1$ \kw{occurs} $\Psi_2$ \kw{occurs within} $[a,b]$
    \begin{center}
    \includegraphics[width=.6\linewidth]{figs/pattern_g}
  \end{center}
The constraint $\Psi_2$ must be satisfied at less one time during $[a,b]$ after $\Psi_1$.

\item $\Psi_1$ \kw{occurs} $n$ \kw{times during} $[a,b]$ \kw{raises} $\Psi_2$
    \begin{center}
    \includegraphics[width=.6\linewidth]{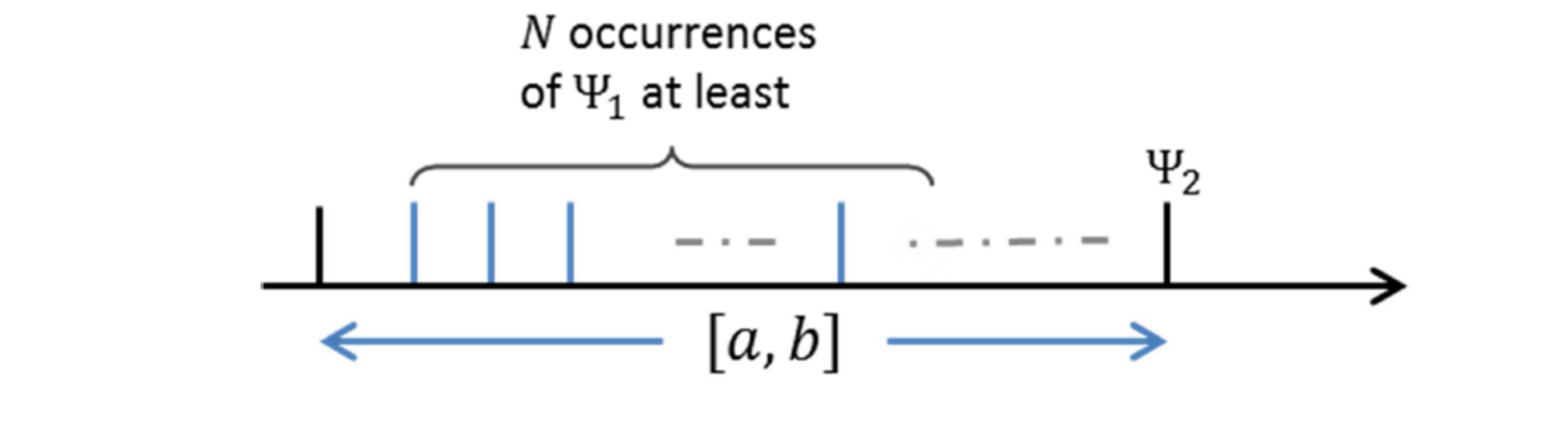}
  \end{center}
When $\Psi_2$ is satisfied at less $n$ times during $[a, b]$, $\Psi_2$ starts to hold at $b$.

\item $\Psi$ \kw{occurs} \kw{at most} $n$ \kw{times during} $[a,b]$
    \begin{center}
    \includegraphics[width=.6\linewidth]{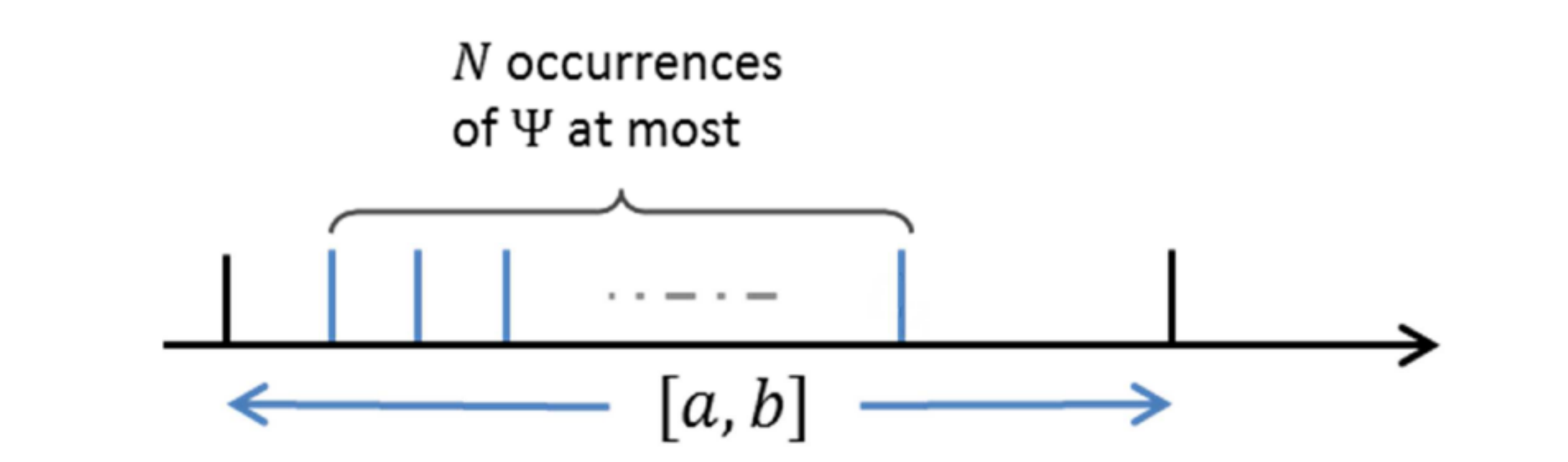}
  \end{center}
As previously mentioned, an occurrence of $\Psi$ is counted when $\Psi$ becomes satisfied. If $\Psi$ holds for
a state in $[a, b]$, to observe $\Psi$ holds for the following one (also in $[a, b]$) does not increase the
occurrence number of $\Psi$.

\item $\Psi_1$ \kw{during} $[a, b]$ {raises} $\Psi_2$
    \begin{center}
      \includegraphics[width=.6\linewidth]{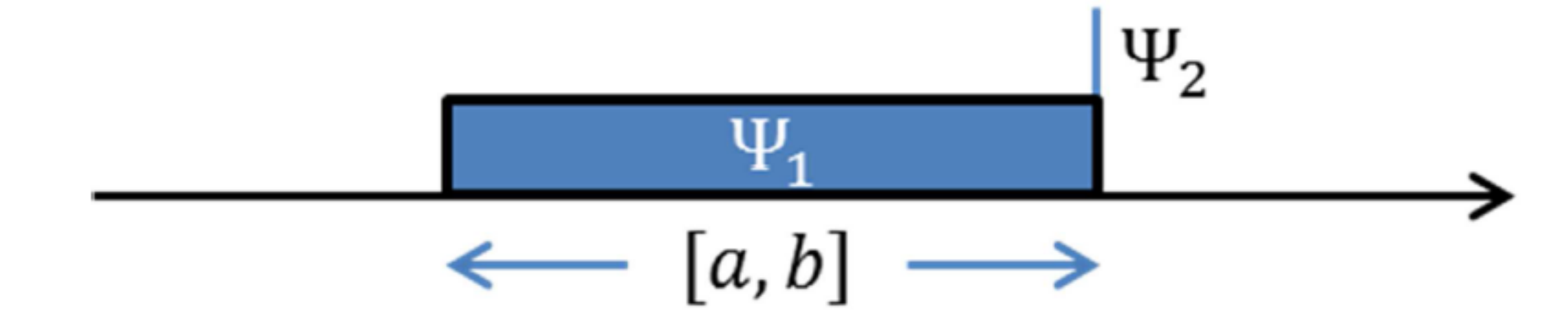}
    \end{center}
If $\Psi_1$ holds during $[a, b]$ then $\Psi_2$ must hold at $b$.

\item $\Psi$ \kw{during} $[a,b]$ \kw{implies} $\Psi_1$ \kw{during} $[a,c]$ \kw{then} $\Psi_2$ \kw{during} $[c, b]$
  \begin{center}
    \includegraphics[width=.6\linewidth]{figs/pattern_k}
  \end{center}
Whenever $\Psi$ holds during $[a, b]$ there exists a split at $c$ of $[a, b]$ such that $\Psi_1$ holds during $[a,c]$ then $\Psi_2$ holds during $[c,b]$.
\end{enumerate}


\newpage

\end{document}